**Title**

- Vortex-free Intrisic Orbital Angular Momentum.
- Hydrodynamic Insights into Photonic Angular Momentum

**Authors**

Wenxiang Yan,[1,2†] Zheng Yuan,[1,2†] Yuan Gao,[1,2] Xian Long,[1,2] Zhi-Cheng Ren,[1,2] Xi-Lin Wang,[1,2] Jianping Ding,[1,2,3,*] and Hui-Tian Wang[1,2,*]

**Affiliations**

[1]National Laboratory of Solid State Microstructures and School of Physics, Nanjing University, Nanjing 210093, China

[2]Collaborative Innovation Center of Advanced Microstructures, Nanjing University, Nanjing 210093, China

[3]Collaborative Innovation Center of Solid-State Lighting and Energy-Saving Electronics, Nanjing University, Nanjing 210093, China

†These authors contributed equally to this work.

*Corresponding author: jpding@nju.edu.cn; htwang@nju.edu.cn

**Abstract**

Optical orbital angular momentum (OAM) has traditionally relied on vortex beams with helical phase fronts imparting quantized intrinsic OAM. Here, we introduce a fundamentally vortex-free framework where intrinsic OAM arises from the natural curvature of light's energy flow—specifically, the caustic geometry of self-accelerating beams whose curved trajectories act as "orbital highways" for photons. This OAM generation mechanism is independent of phase vortices but mirrors celestial orbital motion. Through numerical simulations, experimental characterization, and optomechanical measurements using optical tweezers, we demonstrate intrinsic vortex-free OAM rooted solely in beam intensity architecture. Generalizing beyond geometric caustics to arbitrary optical fields, we demonstrate OAM via curved Poynting-vector energy streamlines, unifying conventional vortex and novel vortex-free OAM under a single quantitative framework. Streamline engineering enables customizable rotational dynamics, including hybrid orbital-cyclonic motions reminiscent of tropical storms, with promising applications in precision optomechanics, optofluidics, and optical analogues of fluid dynamics. This energy-flow perspective offers a versatile platform for designing and quantifying OAM across structured light.

**Teaser**

Energy-flow curvature customizes OAM without vortices, offering direct control over light's rotational dynamics.

**Introduction**

The quest to unravel light's rotational dynamics stretches back over a century. In 1909, Poynting(*1*) first recognized spin angular momentum (SAM) in circularly polarized light—analogous to a spinning top (Figs. 1A and 1D)—and Darwin(*2*) later argued that atomic transitions conserve angular momentum via orbital angular momentum (OAM), though this was confined to rare quantum events. A breakthrough came in 1989 when Coullet et al. drew an analogy between optical vortices and superfluid quantum vortices(*3*), showing that laser cavities could sustain whirlpool-like phase singularities. Building on this, Allen et al. in 1992 demonstrated(*4*) that helical phase fronts, exp($il\varphi$), carry OAM of $l\hbar$ per photon (Fig. 1E), larger than SAM and immediately spurring applications(*5*, *6*) in optical manipulation(*7–9*), metrology(*10*), encryption(*11*), infromation communications(*12*, *13*), and beyond. Consequently, for the past three decades OAM research has concentrated primarily on vortex beams, a focus underscored by the parallel rise in publications on "optical vortices" and "OAM" since 1992 (Fig. S1). Vortex beams, being eigenstates of the OAM operator, provide a complete modal basis in which any arbitrary light field can be represented as their superposition. Therefore, OAM stands as a universal attribute of light fields(*14–16*)—decoupled from the presence of vortices and embodying the intrinsic rotational dynamics of the energy flow. Yet, despite isolated demonstrations of rotational behavior in non-vortex or vortex-synthesis schemes(*17–21*), the broader investigation of intrinsic OAM in vortex-free structured light remains nascent. For instance, self-accelerating beams(*22*)—first predicted by Berry and Balazs in 1979—trace curved "orbital highways" (e.g., parabolic Airy beams(*23*) or spiral solenoids(*24–26*)) that outwardly manifest rich rotational dynamics (Fig. 1F), but systematic exploration of their hidden angular momentum has been lacking.

Here, we introduce a fundamentally new, vortex-free mechanism for generating and tailoring intrinsic OAM in self-accelerating waves and general structured fields. By extending Poynting's mechanical analogy beyond spin(*1*), we show that OAM emerges directly from the curvature of three-dimensional caustics—photon "orbital highways" defined by overlapping geometric rays (Fig. 2)—rather than from helical phase topology. This caustic-curvature–driven OAM in self-accelerating waves, mirroring celestial mechanics (Fig. 1C, F), is rigorously validated through : (i) numerical simulations quantifying OAM via caustic curvature; (ii) precision measurements using momentum-space tomography (Fig. 3); and (iii) direct optomechanical transfer in optical tweezers (Fig. 4). We then generalize beyond geometric optics by mapping exact Poynting-vector streamlines(*27*)—the integral curves of the energy flow—which serve as a complete "roadmap" for photon motion in arbitrary fields. Across both vortex and vortex-free beams, all intrinsic OAM is shown to arise from streamline curvature (Fig. 5). Finally, by tailoring streamline architectures, we enable the active customization of complex rotational dynamics and intrinsic OAM. For instance, by designing fields with intertwined rotational and orbital motions (Fig. 6)—similar to the orbital-cyclonic dynamics of natural phenomena like tropical storms—we construct hybrid beams that combine vortex phases with curved caustic trajectories, giving rise to dual-form intrinsic OAM, with both vortex-based and caustic-induced vortex-free components. The mechanical equivalence of vortex-based and vortex-free intrinsic OAM has been validated in optical tweezers experiments. Together, these results indicate streamline curvature as a unifying, quantitative framework for designing and measuring intrinsic OAM beyond conventional vortex-based approaches, opening new pathways for OAM-related systems including precision optomechanical control, advanced optofluidic devices, and biomimetic rotational systems.

# Results

## 1. Vortex-free Intrinsic OAM in Self-Accelerating Waves

### 1.1 Photonic Orbital Motion: From Mechanical Analogy to Optical Reality

Angular momentum universally characterizes rotational dynamics—from spinning electrons to orbiting galaxies. In classical mechanics, three archetypal forms are identified (Fig. 1A-C): (i) spinning particles (e.g., Earth's spin), (ii) rigid-body rotation (e.g., galactic spirals), and (iii) orbital motion (e.g., planetary orbits). In optics, two forms are well established: SAM in circularly polarized light (Fig. 1D) and vortex-based OAM from helical wavefronts (Fig. 1E). However, the photon analog of orbital motion (Fig. 1C) has long remained elusive despite its evident role in celestial mechanics.

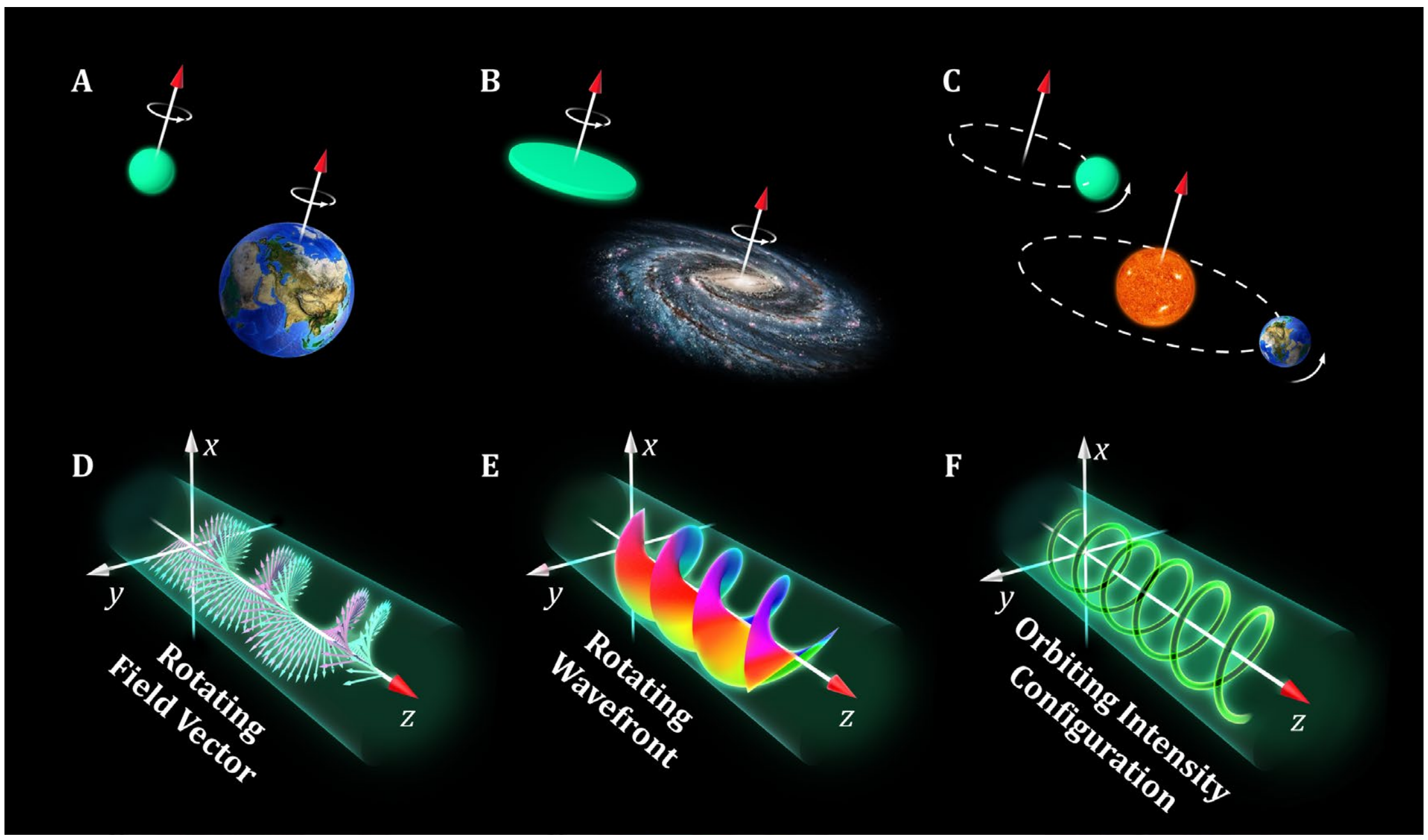

**Fig. 1. Angular momentum models with rotational and orbital dynamics.** In mechanics: (**A-C**) the spinning particle, rigid-body rotation, and the orbital motion of the particle, with corresponding astronomically to astronomical parallels in the Earth's spin, galactic spirals, and planetary orbits (e. g. the Earth's orbit around the Sun), respectively. In optics: (**D**) a circularly polarized beam with rotating electric and magnetic field vectors. (**E**) a vortex beam with the rotating/twisted wavefront. (**F**) vortex-free fields with orbital caustic geometry, e.g. the optical solenoid beams(*25*) with the spiral caustic path in self-accelerating waves. The red arrows depict the angular momentum.

The mechanical momentum and angular momentum of a free particle in Fig. 1C (with the mass *M*) can be derived from its orbital trajectory ( $\boldsymbol{s}(t) = x_m(t)\hat{\boldsymbol{x}} + y_m(t)\hat{\boldsymbol{y}} = r_m(t)\hat{\boldsymbol{r}}$ ):

$$\boldsymbol{p}(t) = M\dot{\boldsymbol{s}}(t) = M(\dot{x}_m(t)\hat{\boldsymbol{x}} + \dot{y}_m(t)\hat{\boldsymbol{y}}) = M[\dot{r}_m(t)\hat{\boldsymbol{r}} + \dot{\varphi}_m(t)r_m(t)\hat{\boldsymbol{\varphi}}] \quad (1)$$

$$J_z(t) = \left[\boldsymbol{s}(t)\times\boldsymbol{p}(t)\right]\cdot\hat{\boldsymbol{z}} = M(x_m(t)\dot{y}_m(t) - y_m(t)\dot{x}_m(t)) = Mr_m^{\ 2}(t)\dot{\varphi}_m(t), \quad (2)$$

where the overdot denotes the first-order derivative to the time $t$ and the subscript "$m$" indicates "mechanical". In our work, we explore the optical analogue of these quantities in vortex-free self-accelerating Bessel-like beams(*23–26*) (Fig. 2), which are characterized by customizable 3D caustic/self-accelerating trajectories ( $\boldsymbol{s}(z) = x_s(z)\hat{\boldsymbol{x}} + y_s(z)\hat{\boldsymbol{y}} = r_s(z)\hat{\boldsymbol{r}}$ ). Specifically, the transverse momentum density and the localized longitudinal OAM (per photon) along these paths are quantified:

$$\boldsymbol{p}_\perp(z) = \hbar k\boldsymbol{s}'(z) = \hbar k(x_s{}'(z)\hat{\boldsymbol{x}} + y_s{}'(z)\hat{\boldsymbol{y}}) = \hbar k[r_s'(z)\hat{\boldsymbol{r}} + \varphi_s'(z)r_s(z)\hat{\boldsymbol{\varphi}}] \quad (3)$$

$$J_{z,local}(z) = \left[\boldsymbol{s}(z)\times \boldsymbol{p}_\perp(z)\right]\cdot\hat{\boldsymbol{z}} = \hbar k(x_s(z)y_s{}'(z) - y_s(z)x_s{}'(z)) = \hbar k r_s{}^2(z)\varphi_s{}'(z), \quad (4)$$

where $k$ represents the free-space wavenumber, the subscripts "$\perp$", "$z$", and "$s$" indicates "transversal", "longitudinal", and "self-accelerating", the single prime symbol denotes the first-order derivative to the variable $z$ in parentheses. Detailed derivations are provided in Supplementary Text 1 with Movie S1. This mechanical-optical correspondence is mathematically exact—the trajectory-dependent angular momentum of a free particle (Eq. 2) directly maps to a photon's OAM along its 3D caustic path (Eq. 4), differing only by the momentum scale ($\hbar k$ versus $M$)). Just as the angular velocity ( $\dot{\varphi}_m(t)$ ) of satellite's orbit determines its angular momentum, the curvature ($\varphi_s'(z)$) of a photon's "orbital highway" (Fig. 1F) governs its OAM magnitude—distinct from vortex-based OAM that relies solely on helical phase fronts. This analogy is further reinforced by the correspondence between the paraxial wave equation in optics and the time-dependent Schrödinger equation in quantum mechanics(*15*, *28*): $-i\partial_z\psi(x,y,z) = (\partial_x^2 + \partial_y^2)\psi(x,y,z)$ governs the paraxial propagation of waves with $z$ = (propagation distance)/$k$; when $z$ is reinterpreted as (time) $\hbar/M$ , it aptly describes the quantum dynamics of a free particle of mass $M$.

Angular momentum is central to understanding rotational and orbital dynamics in both mechanical and optical systems. Rotational or orbital dynamics of energy are also prevalent in vortex-free structured fields(*19*, *23–26*). For example, spirally self-accelerating Bessel-like beams(*24–26*) ((i.e., the optical solenoids in Fig. 1F and Fig. 2E) follow trajectories defined by $\boldsymbol{s}(z)=R_0[\cos(\omega_z z), \sin(\omega_z z)]$ where $R_0$ and $\omega_z$ denotes the spiral radius and the angular velocity. The corresponding transverse momentum density from Eq. 3, expressed as $\hbar k R_0\omega_z[-\sin(\omega_z z), \cos(\omega_z z)]$ (verified in Fig. S3G–I), mirrors the rotational energy dynamics in vortex beams(*4*). The OAM per photon, quantified as $\hbar k(R_0)^2\omega_z$ (Eq. 4), aligns with the mechanical OAM $M(R_0)^2\omega$ (Eq. 2) of a particle in uniform circular motion. In contrast to traditional vortex-based OAM, which relies on rotating wavefronts, this 'vortex-free OAM' arises from orbital caustic curvature, expanding our understanding of optical angular momentum into self-accelerating waves for the first time, and offering novel insights into the interplay between mechanical motion and light dynamics.

### 1.2 Geometric Rendering of OAM Conservation in Structured Light.

Geometric optics(*29*) reveals why usual vortex-free structured beams lack intrinsic OAM (Figs. 2A-D): their caustics ——"photonic energy highways" ——are confined to one-dimensional lines ($\varphi_s(z)$ = 0) or two-dimensional planes ($\varphi_s'(z)$ = 0), which precludes the necessary cross product between transverse momentum and trajectory (Eq. 4). The breakthrough arises with 3D caustic configuration. For instance, spiraling paths with nonzero angular variation or curvature ($\varphi_s'(z) \neq 0$) generate helical momentum components, generating OAM through the same mechanism that governs planetary orbits (Figs. 2E-F). This distinction explains why Airy beams[23] (2D caustics without curvature) carry negligible OAM, whereas 3D caustic beams (e.g., spiral beams) exhibit robust OAM.

Although local OAM along the 3D caustic may vary during propagation (Eq. 4), the global OAM—averaged across the transverse plane—is conserved. This conservation in Bessel-like beams is ensured by the geometric mapping between transverse and longitudinal dimensions, where the integrated energy, momentum, and OAM over the transverse plane equal those summed along the mainlobe (Fig. 2). Consequently, the global OAM per photon and the transverse net momentum(*30*) are expressed as longitudinal integrals along the mainlobe, weighted by a tailored intensity profile $I(z)$:

$$J_z = \langle I(z) J_z(z) \rangle_z / \langle I(z) \rangle_z = \hbar k \left\langle I(z) r_s^{\,2}(z) \varphi_s{}'(z) \right\rangle_z / \langle I(z) \rangle_z , \tag{5}$$

$$\boldsymbol{P}_\perp = \langle I(z) \boldsymbol{p}_\perp(z) \rangle_z = k\left( \left\langle I(z) x_s{}'(z) \right\rangle_z , \left\langle I(z) y_s{}'(z) \right\rangle_z \right), \tag{6}$$

where the shortened notation $\langle\cdot\rangle_z = \int \cdot dz$ denotes longitudinal integration. Detailed derivations are provided in “OAM in Self-Accelerating Bessel-like Beams” of Supplementary Text 1. For clarity, we setting $I(z) = 1$ in the main text. In theory, self-accelerating Bessel-like beams—as with ideal Bessel(*31*) and Airy beams(*22*)—require infinite power and aperture to maintain non-diffracting propagation over an unbounded range ($z \in (-\infty, \infty)$). Finite physical apertures confine these beams to a limited and manageable range ($z \in (a, b)$). Consequently, the expressions in Eqs. 5–6 represent the global OAM and transverse net momentum within the physical aperture, with the right-hand sides corresponding to the z-direction integrals over the effective propagation distance ($z \in (a, b)$). This configuration, based on the geometrical mapping between transverse and longitudinal dimensions, mirrors the interaction between an axicon’s illuminating aperture and its diffraction-free range, encapsulating the overall conservation of angular momentum despite localized dynamic variations.

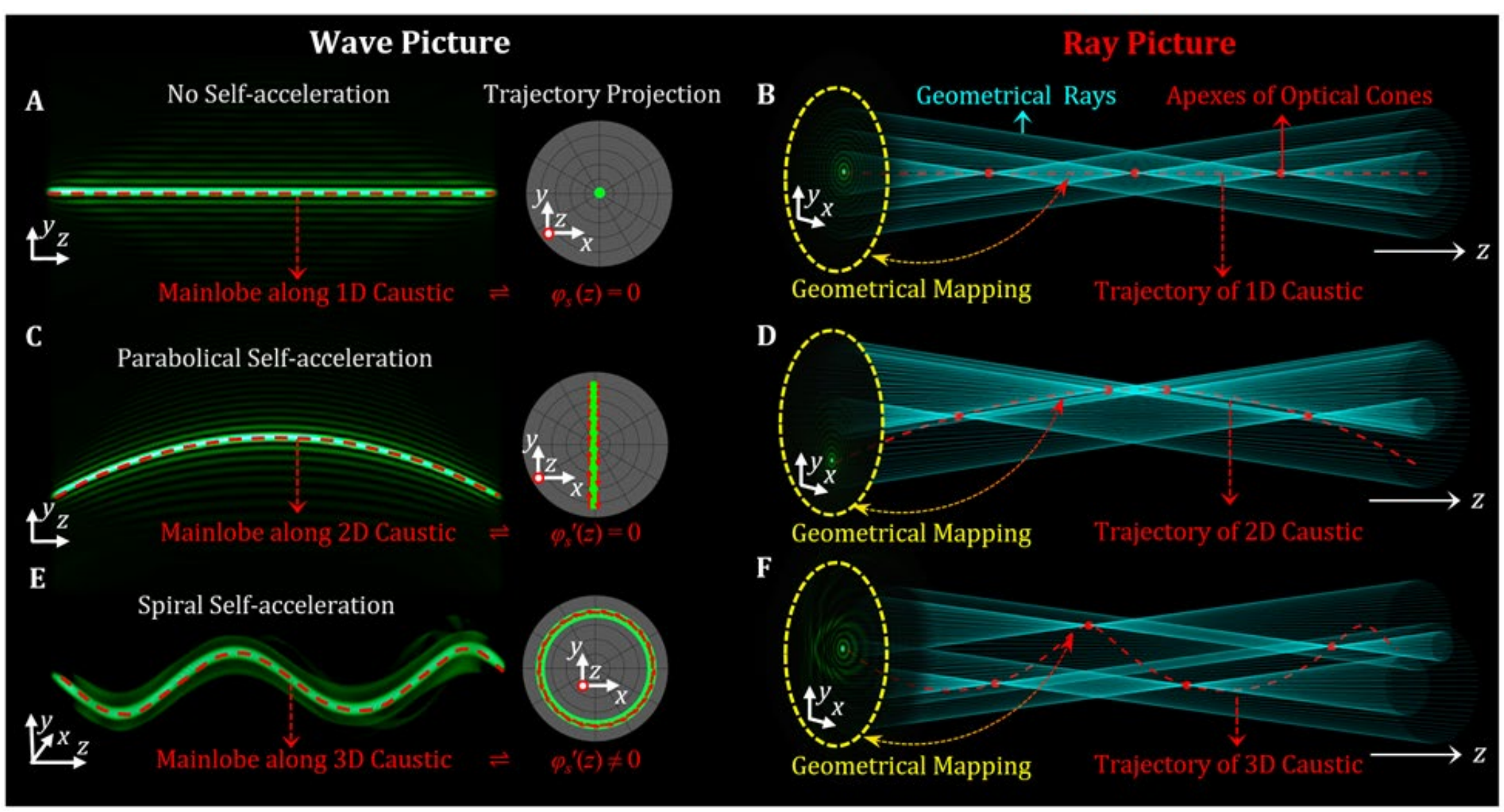

**Fig. 2. Wave and ray pictures of self-accelerating Bessel-like beams.** Subgraphs (**A-B**) feature that the geometrical-optics rays (the cyan lines) of 0-order Bessel beams are lying on coaxial conical surfaces and the caustic (the red-dotted line, acting as orbital highways for

photons)—the straight line (i.e., the mainlobe) constituted by the apexes (the red dots) of optical cones, is termed one-dimensional (1D) caustic. Subgraphs (**C**-**D**) feature that the geometrical-optics rays (the cyan lines) of the parabolically self-accelerating Bessel-like beams(*24*) (like parabolic Airy beams(*23*)) are lying on translational conical surfaces and the parabolic caustic (the red-dotted curve) along the self-accelerating mainlobe, constituted by the apexes (the red dots) of translational light cones, is confined solely to a meridional plane that includes the z axis without curvature, referred to as two-dimensional (2D) caustic. Subgraph (**E**-**F**) feature that the geometrical-optics rays (the cyan lines) of the spirally self-accelerating Bessel-like beams (e.g., optical solenoid beams herein) are lying on translational conical surfaces and the spiral caustic (the red-dotted curve) along the self-accelerating mainlobe, constituted by the apexes (the red dots) of translational light cones, is termed 3D caustic with curvature. The yellow-dashed circles with yellow-red arrows in (**B**), (**D**), and (**F**) indicates geometrical mapping relationship between the transverse and longitudinal dimensions along the caustics that the integration of energy, momentum, and OAM of all light cones across any transverse plane are equivalent to the integration of those quantities along the caustic—the mainlobe constituted by the apexes of all light cones. The insets in the right of (**A**, **C**, **E**) depict 3D trajectory projection onto the *x*-*y* plane with a dot ($\varphi_s(z) = 0$), a line in the radial direction ($\varphi_s'(z) = 0$), and a circle ($\varphi_s'(z) \neq 0$), indicating the 1D, 2D, and 3D caustics, respectively; red arrows depict the transversal momentum along these trajectories by Eq. 3.

### 1.3 Intrinsic Nature: When 3D Caustic Geometry Meets Symmetry.

Physical phenomena are objective and independent of the observer's frame, as exemplified by the translational invariance of intrinsic angular momentum(*32*). Optical SAM is inherently intrinsic and invariant under axis selection, whereas OAM generally comprises both intrinsic and extrinsic components, with the extrinsic part being sensitive to the choice of axis. When the transverse net momentum $\boldsymbol{P}_\perp$ is zero, the extrinsic contribution vanishes, rendering the OAM purely intrinsic and translationally invariant(*30*). See details in "OAM theory of optical fields" of Supplementary Text 1. This condition holds for vortex-based OAM in Bessel and Laguerre–Gauss beams(*30*) and similarly for vortex-free OAM when $\boldsymbol{P}_\perp$— the phasor of $I(z)\boldsymbol{p}_\perp(z)$ along the 3D caustic (Eq. 6) — is a zero vector. With a uniform intensity $I(z) = 1$ over the propagation range $z \in (a, b)$, this requirement reduces to

$$(x_s(a), y_s(a)) = (x_s(b), y_s(b)) \leftrightarrow \boldsymbol{s}(a) = \boldsymbol{s}(b). \tag{7}$$

Equations 5–7 demonstrate that the vortex-free OAM in a uniform-intensity ($I(z) = 1$) self-accelerating beam remains intrinsically translationally invariant if the 3D caustic—the mainlobe—returns to its original transverse position over the effective propagation range (i.e., transversally enclosed trajectories $\boldsymbol{s}(a) = \boldsymbol{s}(b)$ for $z \in (a, b)$). This intrinsic nature arises from the absence of transverse net momentum(*30*). The 3D caustic trajectory itself—$\boldsymbol{s}(z)$ within $z \in (a, b)$, determines the magnitude of intrinsic vortex-free OAM, as described by Eq. 5.

For example, spirally self-accelerating Bessel-like beams (Fig. 2E) that orbit integer cycles over a finite propagation range ($z \in (a, b)$) have spiral caustics that return to their original transverse positions ($\boldsymbol{s}(a) = \boldsymbol{s}(b)$). Under these conditions, the transverse net momentum vanishes, and the global vortex-free OAM remains intrinsically invariant under translational symmetry. Conversely, if the spiral caustic traverses only fractional cycles ($\boldsymbol{s}(a) \neq \boldsymbol{s}(b)$), residual transverse momentum imparts an extrinsic OAM component that depends on the chosen axis, breaking the invariance.

This dynamics parallels vortex beams with fractional topological charges: non-zero transverse net momentum introduces an extrinsic component, disrupting this invariance(*33*). The localized vortex-free OAM along the 3D caustics (Eq. 4) is extrinsic, similar to the extrinsic mechanical angular momentum of an orbiting free particle (Eq. 2). However, the global vortex-free OAM (Eq. 5), remains intrinsic when transverse net momentum (Eq. 6) is absent. This mirrors the behavior of vortex-based OAM: the global OAM in vortex beams is intrinsic, while localized OAM in non-circularly symmetric areas is extrinsic(*30*). This correspondence emphasizes the importance of structural integrity for detecting intrinsic OAM under translational invariance. Our analysis establishes intrinsic vortex-free OAM as a fundamental property of vortex-free structured fields with 3D caustic geometry, e.g., self-accelerating waves.

## 2. Comprehensive Verification of Vortex-Free OAM Dynamics

### 2.1 Quantitative Validation of Vortex-Free OAM.

Our analysis of vortex-free OAM in self-accelerating beams begins with the angular spectrum, characterized by a tailored intensity profile $I(z)$ and arbitrary 3D caustic trajectory of $\boldsymbol{s}(z) = [x_s(z), y_s(z)]$ over the effective propagation distance $z\in (a, b)$, expressed as(*24*)

$$A(k_x,k_y)=\int_a^b \sqrt{I\left(z\right)}e^{ik_x x_s(z)+ik_y y_s(z)}e^{i\sqrt{k^2-q^2}z}e^{ik_z z}dz, \tag{8}$$

where $(k_x, k_y, k_z)$ are the Cartesian wavenumbers, satisfying $k^2 = k_x^2+k_y^2+k_z^2$, and $q$ denotes the transverse wavenumber of constituent plane waves of Bessel-like beams. Figure S2 illustrates a parabolic self-accelerating beam, analogous to parabolic Airy beams. To establish vortex-free OAM as a fundamental property as outlined in Eq. 5, we employed multiple methodologies:

1. Momentum-Space Analysis(*16*) using the angular spectrum,

$$J_z=(\int A^*(\boldsymbol{k}_\perp)\cdot(-i\boldsymbol{k}_\perp\times\frac{\partial}{\partial\boldsymbol{k}_\perp})A(\boldsymbol{k}_\perp)d^2\boldsymbol{k}_\perp)/(i\int A^*(\boldsymbol{k}_\perp)A(\boldsymbol{k}_\perp)d^2\boldsymbol{k}_\perp),\ \boldsymbol{k}_\perp=(k_x,k_y); \tag{9}$$

2. Real-Space Quantification(*16*) via the complex spatial distribution $\psi(x, y, z)$,

$$J_z=(\iint\psi^*(x,y,z)(\boldsymbol{r}\times\nabla)\psi(x,y,z)dxdy)/(i\iint\psi^*(x,y,z)\psi(x,y,z)dxdy; \tag{10}$$

3. Quantitative Experimental Measurement: following the technique of Ref.(*34*) which involves capturing orthogonal Fourier-plane intensity distributions of orthogonal cylindrical lenses and analyzing their first-order moments.

Matlab codes for the first two methods and experimental details for the third are provided in Methods. For localized OAM validation along the 3D caustic (Eq. 4), an aperture pre-truncated the self-accelerating mainlobe in both real-space quantification and experimental measurements. Results from all three approaches consistently corroborate our theoretical predictions (Eqs. 4–5). For clarity, our plots in Fig. 3 directly compare experimental data with theoretical predictions.

Intrinsic vortex-free OAM was first measured in self-accelerating beams with transverse enclosed spiral, trefoil, and 3D parabolic trajectories, which return to their original transverse positions. These configurations are shown in Figs. 3A-3C. Experimental data, encompassing both local and global OAM values, closely match theoretical predictions in Figs. 3D-3F. The calculated intrinsic and extrinsic components, marked in black, highlight the negligible extrinsic contribution for these enclosed trajectories. Further validation from non-uniform intensity profiles ($I(z) \neq 1$) underscores the generality of the framework (Fig. S5, Movie S3). This convergence of theory, computation, and experiment provides important evidence that OAM fundamentally arises from not only phase topology but also 3D caustic geometry.

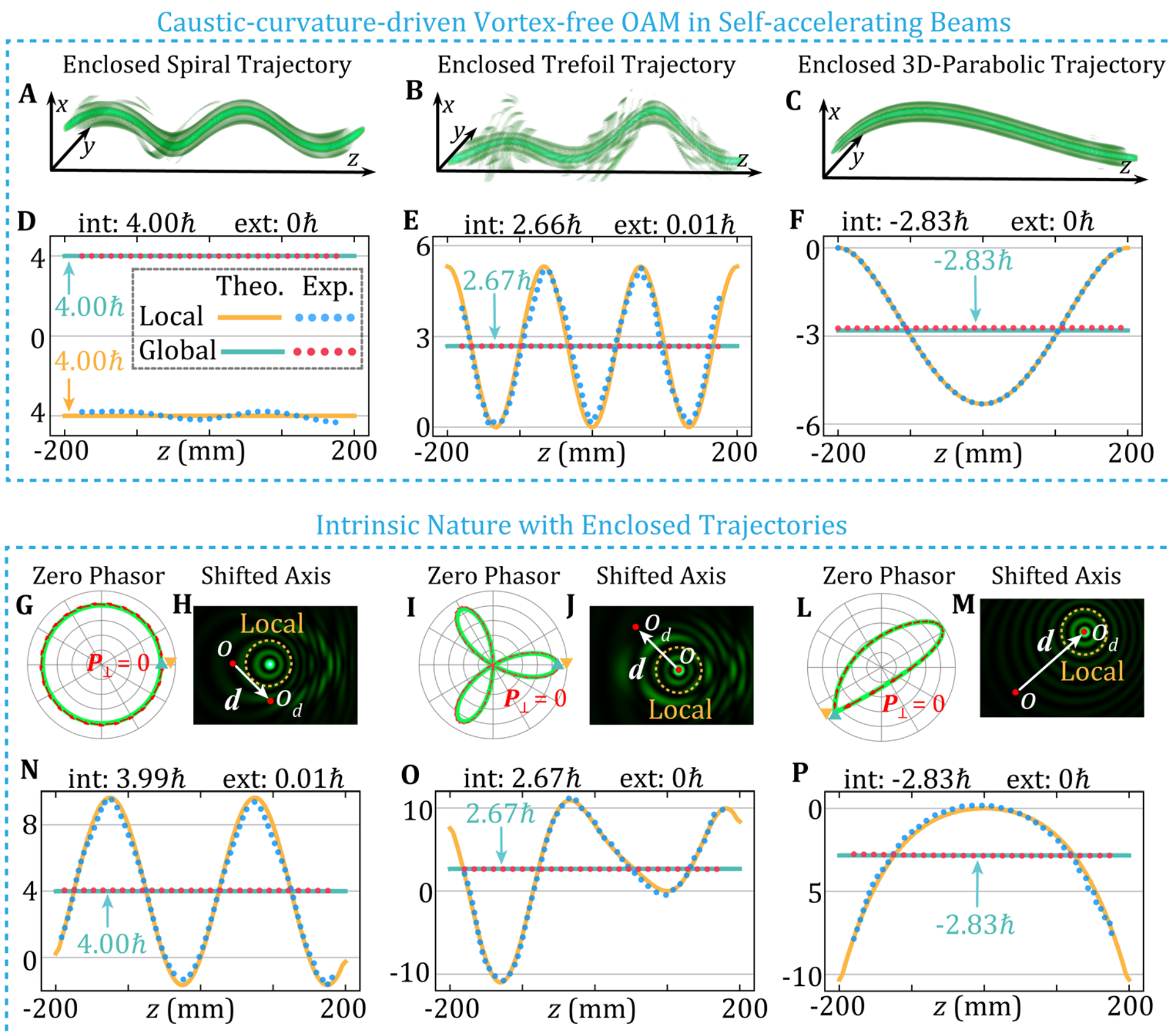

**Fig. 3. Experimental Validation of Intrinsic vortex-free OAM in Self-accelerating beams.** Subgraphs (**A-C**): 3D intensity iso-surfaces for enclosed spiral trajectory $\boldsymbol{s}(z)$=0.1038[cos($2\pi z$/200), sin($2\pi z$/200)] mm, enclosed trefoil trajectory with $\boldsymbol{s}(z)$=0.012[cos($\pi z$/200+$\pi$)+ cos($2\pi z$/200), and enclosed 3D parabolic trajectory with $\boldsymbol{s}(z)$=0.3[(1-$z$/200) (1+$z$/200)$^2$, 1-$z^2$/200$^2$] mm within the range $z\in$ (-200, 200) mm and uniform-intensity $I(z)$ = 1. Subgraphs (**D-F**): Comparison between experimentally measured vortex-free OAM (the dotted curves) and theoretical predictions (the solid curves) from Eqs. 4 and 5. The values at the top indicate the calculated intrinsic (int) and extrinsic (ext) constituents. "Local" refers to local OAM along the 3D caustic/mainlobe, while "Global" represents global OAM across the transverse plane. To avoid aliasing, both ends of the vertical axis in (**D**) are shown as positive. Subgraphs (**G**, **I**, **L**): 3D trajectory projection onto the $x$-$y$ plane, with cyan-up and orange-down triangles marking the start and finish points of the trajectory, respectively. Red arrows depicts the transversal momentum density $\boldsymbol{p}_\perp(z)$ and the phasor is zero ($\boldsymbol{P}_\perp$ = 0) due to the enclosed trajectory. Subgraphs (**H**, **J**, **M**): Intensity map at $z$ = 0 mm, where the brown-dotted circle highlights the local region used for experimental OAM measurements along the mainlobe and the white arrow indicates the translation from the original axis $O$ to the shifted axis $O_d$, with a shifted vector $\boldsymbol{d}$ = (0.1038, -0.1038) mm, $\boldsymbol{d}$ = (-0.12, 0.12) mm, and $\boldsymbol{d}$ = (0.3, 0.3) mm, respectively. Subgraphs (**N-P**): the same as (**D-**

**F**) but for the shifted axis $O_d$. The counter-examples of extrinsic vortex-free OAM with transverse unclosed trajectories are displayed in Fig. S6. Experiment visualizations of these OAM-carrying self-accelerating beams are available in Movie S2.

Translational invariance—the hallmark of intrinsic angular momentum, was validated by shifting the optical axis from $\boldsymbol{O} = (0, 0)$ to $\boldsymbol{O_d} = (x_d, y_d)$, with translation vector $\boldsymbol{d} = (x_d, y_d)$, as shown in Figs. 3H, 3J, and 3M. Figures 3N-3P demonstrate that after translation, the global vortex-free OAM retains intrinsic translational invariance, while the local OAM along the mainlobes shows extrinsic characteristics, mirroring the mechanical angular momentum of an orbiting free particle. Both intrinsic and extrinsic components, marked by numerical values in black, indicate negligible extrinsic contributions, confirming translational invariance. Intrinsic translational invariance of vortex-free OAM potentially allows its application as a novel information carrier in communication systems(*35*, *36*). By leveraging the complexity and unpredictability of optical modulities for vortex-free OAM, potential approaches to optical coding and encryption could be developed(*37*, *38*). Conversely, counter-examples with transverse unclosed trajectories reveal extrinsic vortex-free OAM: unclosed 1.5-cycle spiral and parabolic-linear trajectories exhibit non-zero transverse net momentum (i.e., non-zero Phasor), underscoring the variability in global OAM with axis translation and the breaking of translational invariance, as displayed in Fig. S6.

### 2.2 Mechanical Transfer: From Photonic Orbits to Macroscopic Rotation.

Optical tweezers(*39*), originally proposed by Arthur Ashkin (2018 Nobel Laureate), have become indispensable for manipulating microscopic particles in fields ranging from physics to biology. By transferring angular momentum from light(*9*), this technology enable the rotation of micro- and nano-objects, paving the way for optically driven micromachines, motors, actuators, and advanced biological manipulation. The definitive test of intrinsic angular momentum is its mechanical transfer—a challenge passed by vortex-free OAM.

Figure 4A illustrates our optical manipulation setup, which integrates an inverted confocal microscope (Nikon TE2000-U) with a 4-f complex light field generator featuring a reflective spatial light modulator (SLM, Holoeye Leto). A horizontally polarized, collimated 532 nm laser (Coherent Verdi-v5) is shaped using computer-generated holograms (derived from the angular spectrum in Eq. 8) and focused to create desired vortex-free fields at a power of 90 mW. These fields interact with 3.2 μm-diameter polystyrene microspheres suspended in deionized water within a custom sample chamber—a design featuring an acrylic plate with a through-hole, a glass cover slip at the base, and an open top to minimize boundary resistance, allowing free particle motion at the water's surface.

Consider the spirally self-accelerating beams in Fig. 3A, which exhibit intrinsic vortex-free OAM of $4\hbar$ per photon. Particles are initially drawn toward the beam's mainlobe by the intensity-gradient force, then orbit following this mainlobe until reaching the liquid surface (Fig. 4B). At the surface, the conversion of intrinsic optical angular momentum to mechanical angular momentum induces clockwise rotation (Fig. 4C); reversing the spiral chirality to yield $-4\hbar$ switches the rotation to counterclockwise (Fig. 4D). Similar results are observed with trefoil-shaped self-accelerating Bessel-like beams (Fig. 3B). Visualization of these experiments is available in Movie S4. In contrast, fields lacking intrinsic vortex-free OAM (e.g., a parabolic trajectory in Fig. 2C) or with minimal intrinsic components (e.g., a unclosed parabolic-linear trajectory in Fig. S6) do not induce rotation. *Notably, the particles exhibit solid-core rotation (Fig. 4C-D) due to the absence of a phase* *vortex singularity, thereby enabling compact micromotors.* This optomechanical synergy confirms intrinsic vortex-free OAM, bridging optical and mechanical realms and opening new avenues for optically driven micromachines and biomimetic actuators(*9*) free of phase singularity constraints.

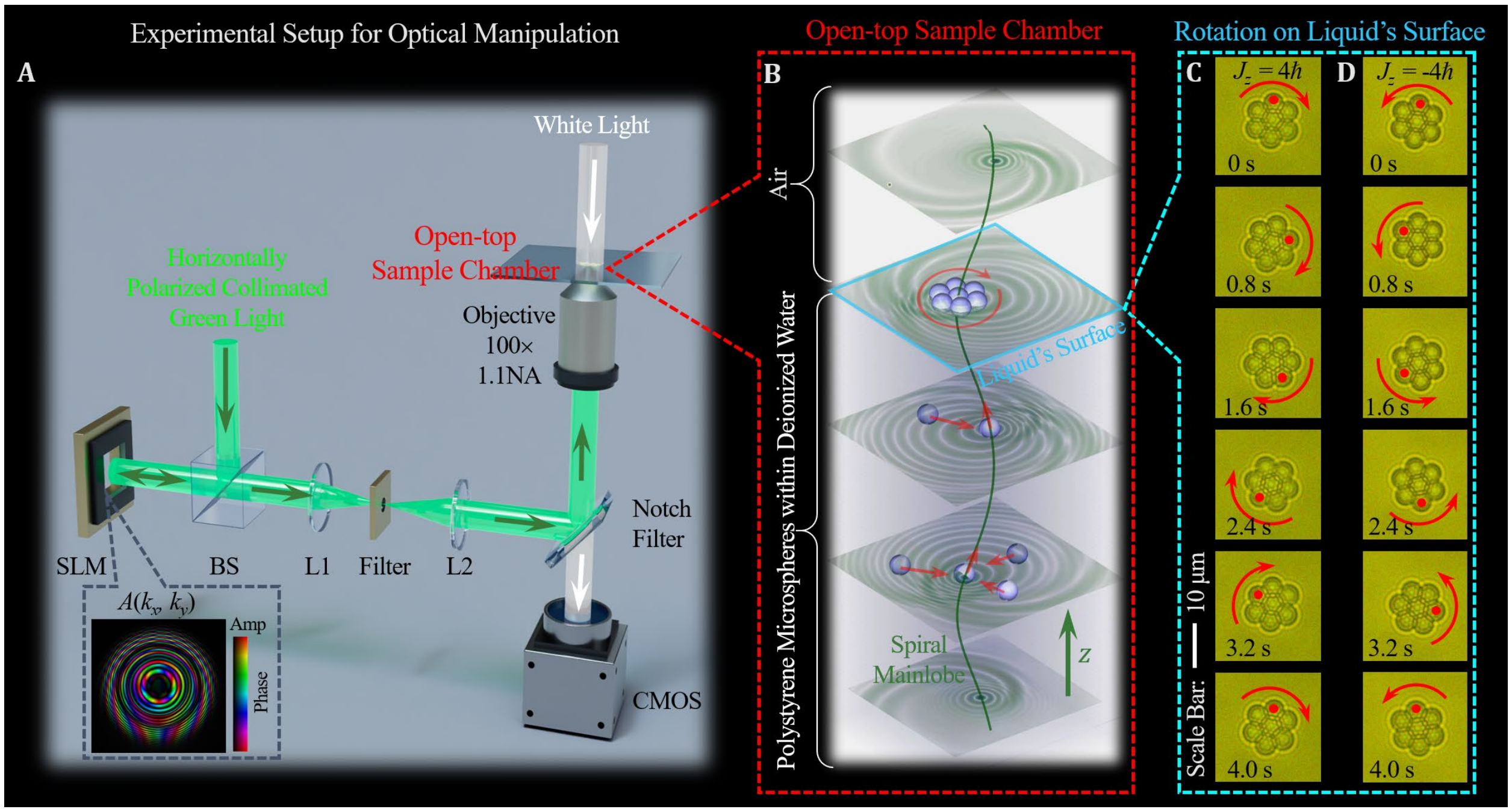

**Fig. 4. Mechanical transfer of vortex-free intrinsic OAM in optical tweezers.** (**A**) Schematic experimental setup for optical manipulation. BS, beam splitters; SLM, phase-only spatial light modulator; L1–2, lenses. The inset depicts the angular spectrum $A(k_x, k_y)$ of a spirally self-accelerating beam from Eq. 8, loaded on SLM after modulation. Subgraph (**B**) depicts the open-top sample chamber with the spirally self-accelerating beam, carrying intrinsic vortex-free OAM of $4\hbar$ per photon, incident from below: the polystyrene microspheres are initially drawn towards the beam's mainlobe and orbit along this mainlobe until reaching the liquid's surface; the particles finally rotate clockwise on the open-top liquid's surface, with the frame images in (**C**). When changing the intrinsic vortex-free OAM from $4\hbar$ to $-4\hbar$ by reversing the chirality of the spiral trajectory, the direction of particle rotation switches to counterclockwise in (**D**). Experiment visualizations are available in Movie S4.

Our theoretical and experimental evidence establishes a vortex-free mechanism for intrinsic OAM in self-accelerating waves, for the first time. This extends the optics-mechanics correspondence: akin to celestial orbits inherently carrying angular momentum through path geometry, structured light could generate intrinsic OAM via caustic curvature—independent of phase vortices. Distinct from yet complementary to conventional vortex-based phase topology, this vortex-free mechanism expands the taxonomy of optical rotational dynamics to encompass three canonical forms (cf. three archetypal mechanical forms in Fig. 1):

(a) SAM in circular polarization (photon spin about its axis);

(b) Vortex-based OAM from helical wavefronts (collective photon rotation about beam axis);

(c) Vortex-free OAM by geometric path curvature (photon orbital motion along curved caustics);

(d) …among other potential forms yet to be fully investigated.

To transcend geometric-optics approximations of caustics, we generalize this vortex-free framework to arbitrary optical fields by extending geometric caustics (the local 'highways' for photon motion) to energy streamlines (the complete 'roadmap' for photon motion)—integral curves of Poynting vectors defining exact photon trajectories(*27*). This yields a unified, streamline-centric methodology to analyze and customize complex rotational dynamics, as demonstrated next.

## 3. Universal Vortex-free Framework for Analyze and Customize Rotational Dynamics

### 3.1 Unifying Analysis of Rotational Dynamics through Energy Streamlines.

The analogy between the dynamics of lasers and fluid/superfluids systems traces back to the early 1970s, when laser physics equations were reduced into the complex Ginzburg–Landau equations(*40*). This hydrodynamic framework(*41*) has since provided critical insights into superconductivity, superfluidity, and Bose-Einstein condensation, catalyzing deeper explorations of hydrodynamic phenomena in optical fields(*42–44*)—including chaos, multistability, and turbulence, all theoretically predicted and experimentally observed in laser systems(*45–47*). Notably, Coullet et al., formally established the concept of optical vortices in 1989 by drawing direct parallels to hydrodynamic vortices(*3*). Building on this foundational understanding, we adopt a hydrodynamic approach to reformulate OAM analysis through energy streamlines(*27*, *48*) — integral curves of the Poynting vector. For a scalar wavefield $\psi(\boldsymbol{R})$ with $\boldsymbol{R} = \{x, y, z\}$, the Poynting vector $p(\boldsymbol{R})$ (representing the expectation value of the local momentum operator(*27*)), is given by:

$$\boldsymbol{p}(\boldsymbol{R}) = \mathrm{Im}\,\psi^*(\boldsymbol{R})\nabla\psi(\boldsymbol{R}) = |\psi(\boldsymbol{R})|^2\,\nabla\arg\psi(\boldsymbol{R}), \tag{11}$$

This expression constitutes a fundamental component for quantifying optical OAM(*16*). Energy streamline trajectories $\boldsymbol{R}(z) = \{r(z), \varphi(z), z\}$ are derived by solving the hydrodynamic differential equations:

$$dr(z)/dz = p_r(\boldsymbol{R}(z))/p_z(\boldsymbol{R}(z)), \quad d\varphi(z)/dz = p_\varphi(\boldsymbol{R}(z))/\left[r(z)p_z(\boldsymbol{R}(z))\right]; \tag{12}$$

where $\boldsymbol{p} = \{p_x, p_y, p_z\} = \{p_r, p_\varphi, p_z\}$. These energy streamlines, often likened to "Bohmian trajectories" of light propagation, represent experimentally measurable paths of average photon trajectories(*27*, *49–51*). In quantum physics, the trajectories of the Poynting vector in light (or quantum-mechanical waves) are described as streamlines in the Madelung hydrodynamic interpretation(*52*), which are later regarded as paths of quantum particles in the Bohm–de Broglie interpretation(*53*, *54*).

Mapping these energy streamlines—the complete 'roadmap' for photon motion—reveals the underlying rotational dynamics of arbitrary optical fields, where angular momentum universally originates from streamline curvature (Fig. 5). This hydrodynamic framework unifies diverse canonical forms of optical rotational dynamics:

(I) **Conventional vortex beams** with intrinsic OAM (Figs. 5A–B) exhibit helical streamlines with non-zero curvature ($\varphi'(z) \neq 0$), driving collective photon rotation about the beam axis—analogous to fluid spiraling down a drain;

(II) **Twisted elliptical Gaussian beams**(*19*) with intrinsic OAM (Figs. 5C) generate tilted streamlines with non-zero curvature ($\varphi'(z) \neq 0$), inducing torsive photon motion;

(III) **Self-accelerating beams** with intrinsic OAM (Figs. 5D–E) feature orbital streamlines with non-zero curvature ($\varphi'(z) \neq 0$), enabling photon orbital motion along curved caustics that mimics planetary orbits.

(IV) **Beams devoid of OAM** (Figs. 5F–G) maintain streamlines with zero curvature ($\varphi'(z) = 0$), precluding angular momentum generation.

Within this unified hydrodynamic perspective, each energy streamline embodies a 3D orbital configuration of localized energy, governed by the trajectory $\boldsymbol{R}(z)= \{r(z), \varphi(z), z\}$. Applying the mechanical analogy established in Eqs. (4), the vortex-free OAM of individual streamlines is quantified by their curvature, while the global OAM of arbitrary optical fields derives from the weighted average of OAM contributions across all streamlines, thereby generalizing the framework beyond self-accelerating waves to encompass arbitrary structured light.

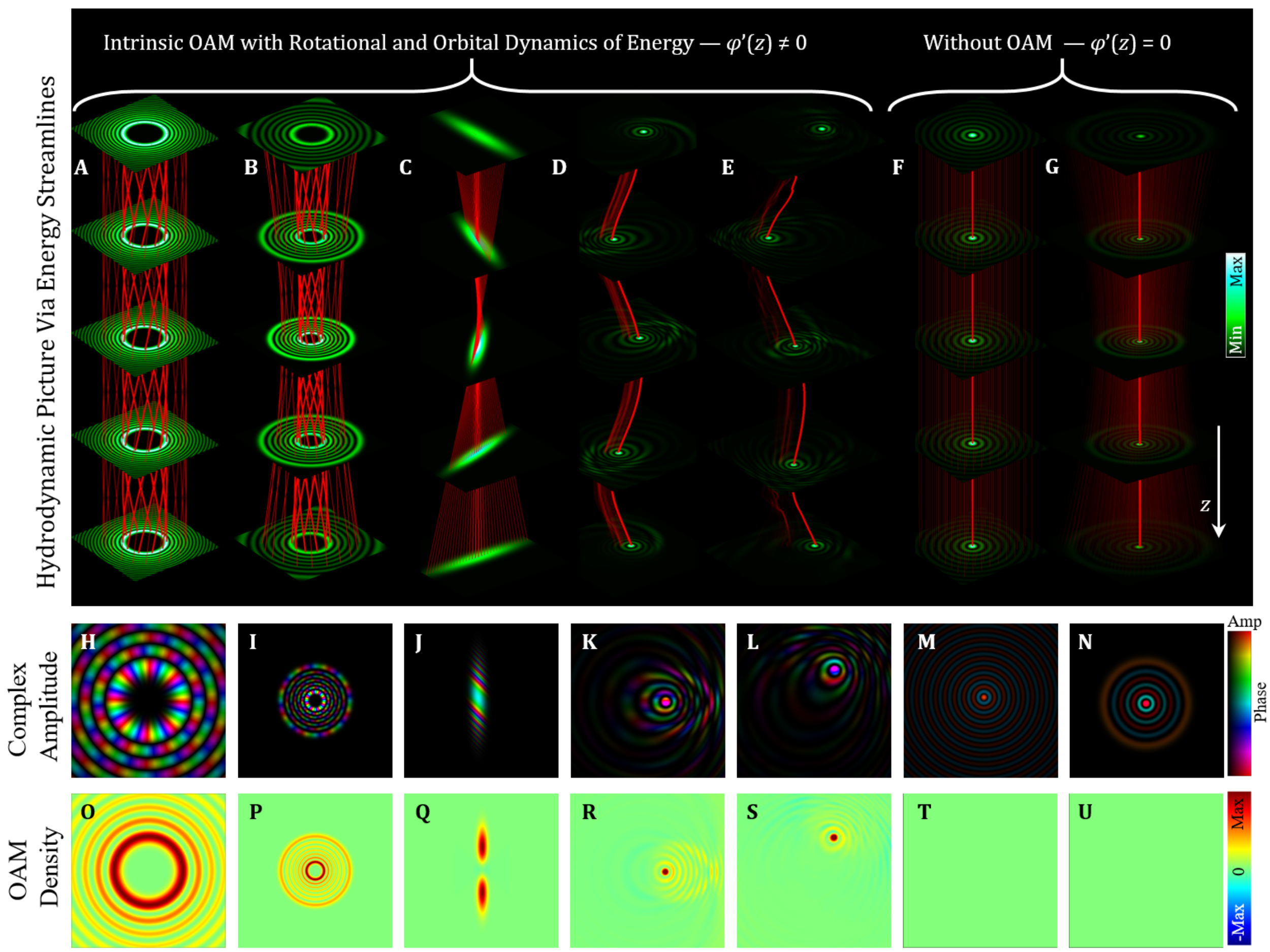

**Fig. 5. Hydrodynamic picture via energy streamlines.** The helical energy streamlines, depicted by the red curves, in (**A**) high-order Bessel beams and (**B**) Laguerre-Gaussian beams indicate the rotational dynamics of energy. The rotational energy streamlines of (**C**) an elliptical Gaussian beam focussed by a tilted cylindrical lens(*19*) indicate the orbital dynamics of energy. The orbiting energy streamlines along the mainlobes of spirally (**D**) and trefoilly (**E**) self-accelerating Bessel-like beams in Figs. 3(A) and (F), indicate the orbital dynamics of energy in vortex-free fields. For the (**F**) Bessel beam and (**G**) Laguerre-Gaussian beam with $l = 0$, without any spiral phase and 3D orbital intensity configuration, the streamlines exhibits neither rotational nor orbital dynamics of energy with $\varphi'(z) = 0$, and consequently precludes OAM. The streamlines herein are directly drew from the “streamline” function with the distributions of Poynting vector caculated in Matlab, which are consistent with the analytical solutions from the hydrodynamic differential equations in this work. (**H**-**N**) The corresponding complex amplitude distributions and (**O**-**U**) OAM density distributions at $z = 0$. The corresponding OAM density distributions upon propagation are provided in Movie S5.

For instance, the energy streamlines of Bessel beams (Fig. 5A) follow helical trajectories(*27*) described by $\boldsymbol{R}(z)=\left\{r_0,\varphi_0+lz\big/\sqrt{k^2-q^2}r_0^{\,2},z\right\}$, where $r_0$ and $\varphi_0$ is the initial position of each streamline. Substituting $\boldsymbol{R}(z)$ into the mechanical analogy (Eq. 4) reveals that the OAM per photon for each helical streamline geometry — consistently equals $l\hbar$. Consequently, the global OAM of Bessel beams within this streamline-centric framework, derived from averaging contributions across all streamlines, yields $l\hbar$ per photon—exactly matching electrodynamic predictions in Eq. 10. Detailed analyses for these optical modalities in Fig. 5 (including Laguerre-Gaussian beams, twisted elliptical Gaussian beams, self-accelerating waves) are provided in the Supplementary Text 2, unanimously confirming consistency with electrodynamic theory. Crucially, this methodology aligns with foundational electrodynamic definitions of optical OAM(*14*, *16*), ensuring universality and broad applicability (Supplementary Text 3). By extending from caustic trajectories (local "orbital highways") to energy streamlines (complete "roadmap" for photon motion), this framework unifies the analysis of rotational dynamics and OAM quantification across optical modalities—including both vortex-based and vortex-free fields—through measurable streamline geometry. Most significantly, it enables customizable design of complex rotational dynamics via streamline engineering—a capability we demonstrate next.

### 3.2 Customizing Rotational Dynamics through Streamline Engineering.

Natural rotational dynamics exhibit inherent complexity and hybridization, as exemplified by Earth's spin-orbital coupling and tropical cyclones, which undergo collective orbital motion while simultaneously rotating locally. These phenomena transcend isolated mechanical modalities illustrated in Figs. 1A–1C (spin, rigid-body rotation, orbital motion). Through momentum-space angular-spectrum engineering, we sculpt the beam's momentum/Poynting-vector field and precisely steer its energy streamlines via hydrodynamic differential equations (Supplementary Text 4). This enables deterministic customization of complex rotational dynamics in light. To emulate intertwined dynamics such as orbiting hurricanes, we strategically configure Cartesian energy streamlines as:

$$\boldsymbol{R}(z)=\left\{x_s(z)+r_0\cos(\varphi(z)),y_s(z)+r_0\sin(\varphi(z)),z\right\},\ where\ \varphi(z)=\varphi_0+\frac{l}{r_0^{\,2}\sqrt{k^2-q^2}}z. \tag{13}$$

This equation synthesizes the combined dynamics of Bessel vortex beams $\boldsymbol{R}(z)=\left\{r_0,\varphi_0+lz\big/\sqrt{k^2-q^2}r_0^{\,2},z\right\}$, and vortex-free self-accelerating beams $\boldsymbol{R}(z) = \{x_s(z), y_s(z), z\}$ (Supplementary Text 2). Following momentum-field configuration and angular-spectrum engineering (Supplementary Text 4), the constructed beam in real-space yields an annular vortex mainlobe that follows the predefined self-accelerating trajectory $\mathbf{s}(z) = [x_s(z), y_s(z)]$ (Figs. 6A-C), and exhibits the targeted cyclone-like rotational and orbital motion. Geometrical-optics analysis reveals that skew rays align on translational hyperboloidal surfaces, generating 3D tubular caustics along the annular vortex mainlobe (Figs. 6D-F, red dotted curves). The integrated results in Figs.

6G-I confirm the coexistence of hybrid rotational-orbital dynamics, in agreement with the pre-designed streamline trajectories (Eq. 13).

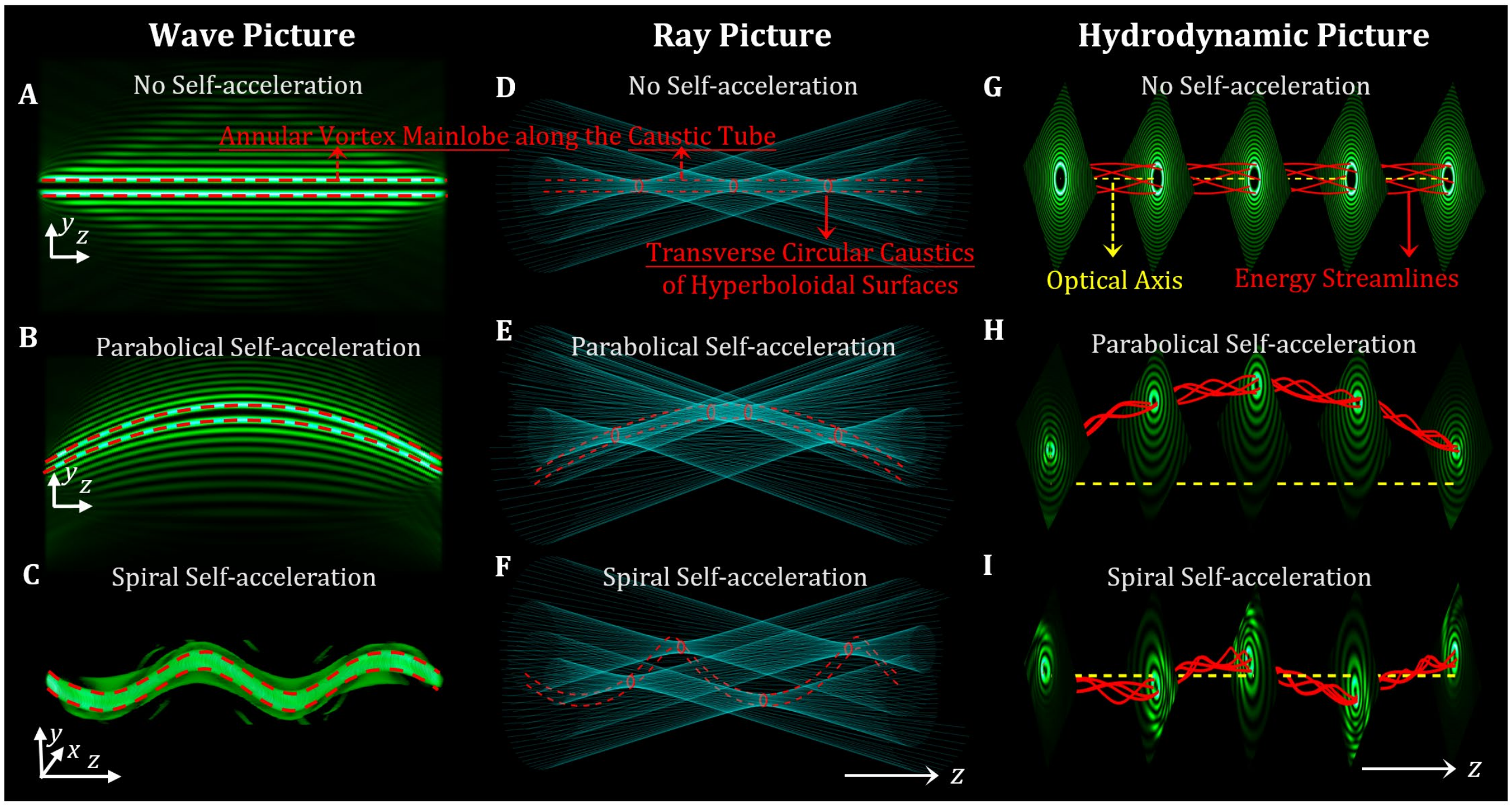

**Fig. 6. Wave, ray and hydrodynamic pictures of hybrid rotational-orbital dynamics.** Subgraphs (**A-C**) features vortex beams with no, parabolical, and spiral self-acceleration, and the red-dotted lines depicts the trajectories of annular vortex mainlobes. Subgraphs (**D-F**) features that the geometrical-optics rays (the cyan lines) of self-accelerating vortex beams in (**A-C**) are lying on hyperboloidal surfaces and the caustic "tubes" (the red-dotted lines) are constituted by the transverse circular caustics (red circles) of hyperboloidal surfaces. Subgraphs (**G-I**) features that by conceptualizing optical fields as energy streamlines (red curves), this hydrodynamic framework clarifies the intertwined rotational and orbital dynamics within the caustic "tubes" (annular vortex mainlobes) of self-accelerating vortex beams. The yellow-dashed lines indicate the optical axis. The streamlines are directly drew from the "streamline" function with the distributions of Poynting vector calculated in Matlab from the hydrodynamic differential equations, which are consistent with the pre-design of Eq. 13.

The global OAM, quantified via streamline curvature in Eq. (13) and derived from Eq. (5), is expressed as:

$$J_z = \hbar k \left\langle I(z)(x_s(z)y_s'(z) - y_s(z)x_s'(z)) \right\rangle_z / \left\langle I(z) \right\rangle_z + l\hbar. \quad (14)$$

Here, the first term corresponds to vortex-free OAM (Eq. 5, Cartesian coordinates), arising from collective orbital motion of self-accelerating trajectories, and the second term is vortex-based OAM, originating from local vortex-induced rotation. This unifies within the streamline-centric framework that synthesizes intertwined rotational-orbital dynamics. Crucially, momentum analysis confirms that the vortex-based component is inherently intrinsic, while the vortex-free component attains intrinsic properties only under transversally enclosed trajectories ($\boldsymbol{s}(a) = \boldsymbol{s}(b)$ for $z \in (a, b)$ per Eq. (7), which eliminates transverse net momentum. This intrinsic hybrid OAM aligns with electrodynamic theory and is rigorously validated through analytical derivations (Supplementary

Text 5), experimental measurements (Supplementary Text 6 with Movie S6), and optical tweezers experiments (Movie S7; setup in Fig. 4A). The latter (optical tweezers experiments) demonstrates mechanical equivalence between both OAM forms: tunable particle rotation states directly correlate with hybrid OAM values, confirming their shared dependence on streamline geometry.

Building on the established mechanical equivalence(*55*) between SAM and vortex-based OAM, we engineered spirally self-accelerating vortex beams (Fig. 6C) with two components: a static vortex-free OAM ($2\hbar$ per photon) maintaining time-invariant clockwise spiral trajectories; a dynamically tuned vortex-based OAM modulating topological charges ($l$ = 4, 2, 0, −2, −4). This configuration yields hybrid OAM values of $(l+2)\hbar$ per photon. Experimentally, dynamic tuning of $l$ while fixing vortex-free OAM induces predictable microparticle rotation transitions (Movie S7): *high-speed clockwise* at $6\hbar$, *mid-speed clockwise* at $4\hbar$, *low-speed clockwise* at $2\hbar$, *critical near-static* at $0\hbar$, and *direction-switched counter rotation* at $-2\hbar$. The monotonic OAM-velocity relationship validates mechanical equivalence, proving both forms depend fundamentally on streamline geometry. Future work will extend this framework to SAM–vortex-free OAM interactions via birefringent particle manipulation(*55*) with circularly polarized vortex-free beams.

## Discussion

Our work revisits the foundational concept of orbital angular momentum (OAM), which has traditionally been tied to phase vortices in structured light. By moving beyond vortex-based eigenstates, we reveal that OAM is an intrinsic feature of energy-flow dynamics in any light field. Specifically, we demonstrate that self-accelerating waves acquire “vortex-free” OAM through the curvature of three-dimensional caustic trajectories—photon “orbital highways”—rather than through twisted phase singularities (Fig. 2). Numerical simulations, high-precision measurements (Fig. 3), and direct mechanical transfer in optical tweezers experiments (Fig. 4) consistently show that this caustic-curvature-induced angular momentum is universal and independent of phase vortices, mirroring the mechanical angular momentum (Fig. 1).

Extending beyond the geometric-optics picture, we replace caustic highways with complete energy streamlines, the integral curves of the Poynting vector, to capture underlying rotational dynamics in arbitrary optical fields (Fig. 5). In this unified, streamline-centric framework, OAM emerges naturally from energy-flow curvature, offering a general method to analyze and customize complex light motions—including hybrid rotational-orbital dynamics akin to orbiting tornadoes (Fig. 6). The mechanical equivalence of vortex-based and vortex-free intrinsic OAM has been validated in optical tweezers experiments. (Deeper explorations on tailoring light’s dynamics via streamline engineering are detailed in Ref. (*56*), beyond our present scope). These results establish energy-flow curvature as a complementary mechanism beyond phase topology for controlling intrinsic angular momentum.

By decoupling OAM from vortices, our theory may open the door to tailored rotational dynamics in diverse platforms. Because energy streamlines directly reveal the direction of optical forces(*27*) and possess compact vortex/singularity-free rotation-driving characteristics (Fig. 4), we anticipate that this vortex-free energy-flow-centric perspective will inform the development of precision optomechanical control(*57*), advanced optofluidic manipulations(*9*), and innovative fluid-dynamic analogues in optics(*58*). Moreover, the intrinsic invariance of vortex-free OAM potentially allows its application as a novel information carrier in communication systems(*35*, *36*). By leveraging the complexity and unpredictability of complex vortex-free optical modulities, we expect that potential approaches to optical coding and encryption technologies could be developed(*37*, *38*).

## Materials and Methods

### Measurement of Vortex-free OAM

The Matlab codes for numerically computing OAM of vortex-free self-accelerating beams have integrated to the repository of Github at https://github.com/WenxiangYan/OAM. According to the method proposed in ref.(*34*), the OAM can be measured using the equation:

$$J_z \approx \frac{\hbar k}{f_c}(\iint_\infty x\xi I_{yf}(x,\xi)dxd\xi - \iint_\infty \eta y I_{xf}(\eta,y)d\eta dy)/(\iint_\infty I_{xf}(\eta,y)dxdy), \quad (15)$$

where $f_c$ is the focal length of a pair of cylindrical lenses with perpendicular focusing dimensions (i.e., the $x$- and $y$-dimensions). $I_{xf}(\eta, y)$ and $I_{yf}(x, \xi)$ are the intensity distributions of the focal planes of the cylindrical lenses focusing along the $x$ and $y$ dimensions, respectively. The coordinates of these planes are $(\eta, y)$ and $(x, \xi)$ , which can be measured by CMOS2 and CMOS3 as shown in Fig. 7.

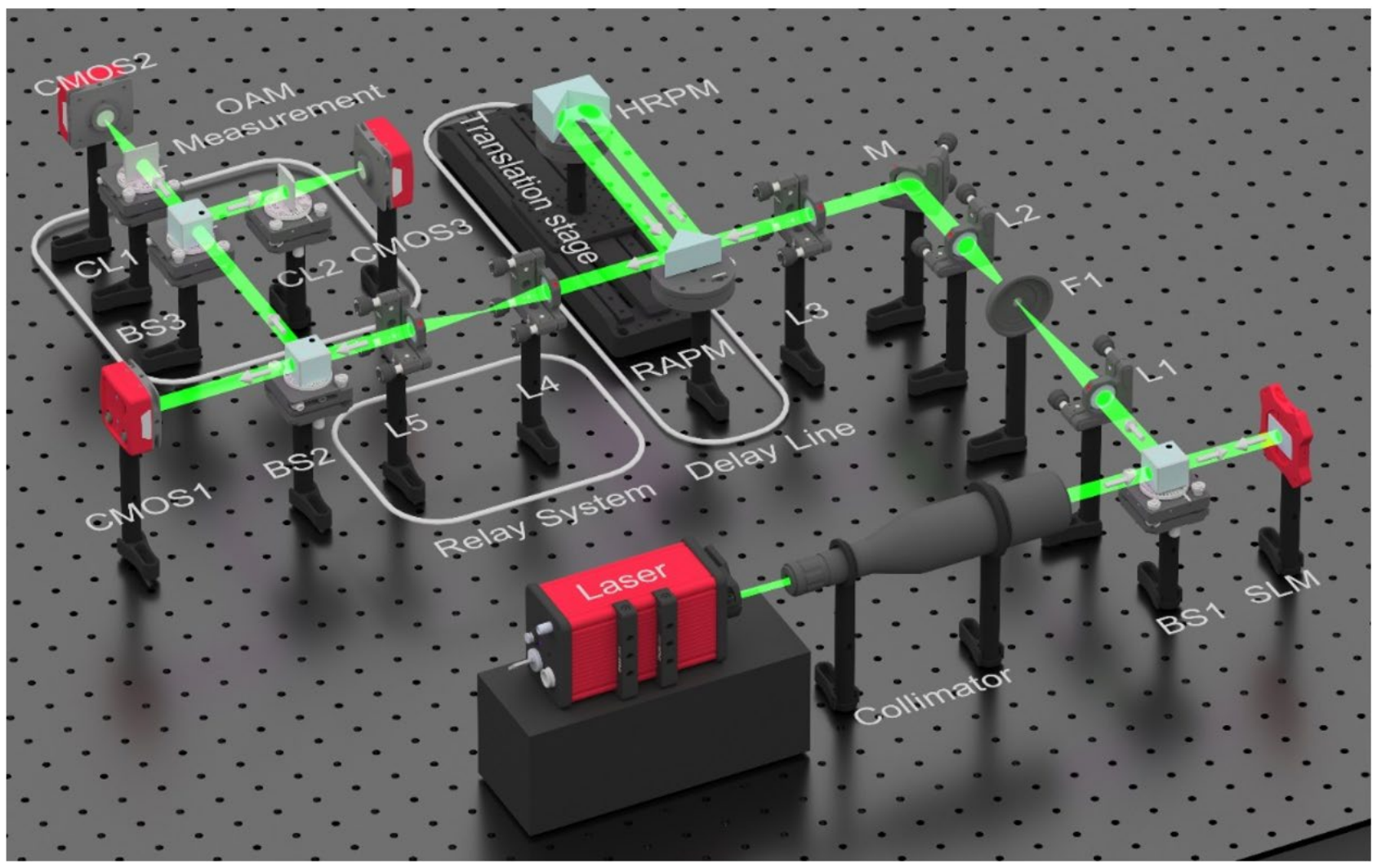

**Fig. 7**. Schematic experimental setup for measuring the optical OAM. BS1-3, beam splitters; SLM, phase-only spatial light modulator; L1–5, lenses; F1, filter; M, mirror; RAPM, right-angle prism mirror; HRPM, hollow roof prism mirror from LBTEK; CMOS1-3, complementary metal-oxide semiconductor cameras; CL1-2 cylindrical lenses. Actual experimental setup is displayed in Fig. S8.

***Generation.*** The experiment setup is shown in Fig. 7: A reflective SLM (Holoeye GAEA-2), imprinted with computer-generated hologram patterns (the Fourier marks), transforms a collimated laser light wave into a complex field corresponding to the angular spectrum of self-accelerating Bessel-like beams in real-space coordinates, with the help of spatial filtering via a 4-F system consisting of lenses L1 and L2, and a filter F1 as well. The resulting field is responsible for generating self-accelerating Bessel-like beams in the focal volume of lens L3.

***Detection.*** A delay line, consisting of right-angle and hollow-roof prism mirrors and a translation stage, enables the different cross-sections of self-accelerating Bessel-like beams to be imaged on a complementary metal-oxide semiconductor camera CMOS1 (Dhyana 400BSI from LBTEK) after passing through a relay 4-F system consisting of two lenses. The combination of the delay line and the relay system enables the recording of intensity cross-sections at different z-axial locations relative to the focal plane of lens L3.

***Global OAM Measurement.*** Beam splitter BS2 divides the beam, directing one part to CMOS1 for imaging and the other to a pair of cylindrical lenses for OAM measurement. The optical field at the focal plane of L5 is focused onto CMOS2 and CMOS3 (PCO. edge 4.2bi), enabling the calculation of global OAM using Eq. 15. The combination of the delay line, the relay system, and the OAM measurement section work together to measure global OAM at different z-axial locations.

***Local OAM Measurement.*** A filter, fabricated with photoetched chrome hole patterns on a glass substrate, is inserted at the front focal plane of L4 to intercept the self-accelerating mainlobe (Fig. S8). Since each photon in the non-diffracting, self-accelerating structure around the mainlobe carries the same OAM per photon ($\hbar k(x_s(z)y_s'(z)-y_s(z)x_s'(z))$, Eq. 4), the local area can be selected with some robustness (e.g., the mainlobe with several sidelobes) without affecting the measured OAM. The apertured areas are transversely synchronized with the self-accelerating mainlobes as the z-axial locations change by the delay line, monitored by CMOS1.

***Translated Axis.*** Translational invariance of intrinsic angular momentum was validated by shifting the optical axis from $\boldsymbol{O} = (0, 0)$ to $\boldsymbol{O_d} = (x_d, y_d)$, with translation vector $\boldsymbol{d} = (x_d, y_d)$, as shown in Figs. 3H, 3J, and 3M. In the experiments, the actual optical axis remained fixed while the beam was translated by $-\boldsymbol{d} = (-x_d, -y_d)$. Following the Fourier phase-shifting theorem, the angular spectrum of the translated beam is expressed as $A_d(k_x, k_y) = A(k_x, k_y)\exp(-ik_x x_d - k_y y_d)$.

## Acknowledgments

**Funding:**

National Key Research and Development Program of China (2023YFA1406903), National Natural Science Foundation of China (12374307, 12234009, 12427808)

**Author contributions:**

Conceptualization: W.Y. and J.D.
Methodology: W.Y., and J.D.
Investigation: W.Y. and J.D.
Experiments: W.Y. and Z. Y.
Visualization: W.Y.
Supervision: J.D. and H.-T.W.
Writing—original draft: W.Y., J. D.and H.-T.W.
Writing—review & editing: All authors.

**Competing interests:** The authors declare no competing interests.

**Data and materials availability:** All data that support the findings of this study are available in the main text and the supplementary materials, or available from the corresponding author on reasonable request.